\documentclass[12pt]{iopart}
\usepackage[T1]{fontenc}		
\usepackage{graphicx}	
\usepackage[section]{placeins}	
\usepackage{ae}					
\begin{document}

\def\ii{\mathrm{i}}
\def\ee{\mathrm{e}}

\title{High order non-unitary split-step decomposition of unitary operators}

\author{Toma\v z Prosen and Iztok Pi\v zorn}

\address{Physics Department, Faculty of Mathematics and Physics, University of Ljubljana,
Jadranska 19, SI-1000 Ljubljana, Slovenia}

\begin{abstract}
We propose a high order numerical decomposition of exponentials of hermitean 
operators in terms of a product of exponentials of simple terms, following an 
idea which has been 
pioneered by  M. Suzuki, however implementing it for complex coefficients.
We outline a convenient fourth order formula which can be written compactly for 
arbitrary number of noncommuting terms in the Hamiltonian and which is superiour to the
optimal formula with real coefficients, both in complexity and accuracy.
We show asymptotic stability of our method for sufficiently small time step 
and demonstrate its efficiency and accuracy in different numerical models.
\end{abstract}

\section{Introduction}
While exponentials of operators are very common in every field of 
quantum physics, but also in classical physics, their evaluation is nevertheless
numerically a very demanding operation. 
For example, in quantum physics, this task usually emerges when one wants to compute
a time-evolution, either in real time, for example when computing dynamical correlations, or
in imaginary time, when computing thermodynamic averages like in quantum Monte Carlo 
simulations. A similar decomposition of classical time evolution, which can also be 
interpreted in terms of unitary operators, is known as symplectic integration.
 
For an operator which can be written as a sum of several parts of 
which exponential operators are exactly determinable, the well known 
Suzuki-Trotter\cite{trotter1059,suzuki5176,suzuki2685,suzuki16592,suzukiJPhSocJ61,suzukiPLA146,suzukiJMathPh32} 
decomposition scheme can be used. The operator $e^{\ii z \sum_j A_j}$ is approximated
by a product of operators $e^{\ii z p_{k_j} A_j}$ with {\em real} coefficients $p_k$ such that the desired order of accuracy is achieved. 
We will show in the present paper that following the same principles but not restricting 
to real coefficients the same order can be achieved using a smaller number of factors. 
Furthermore, the order of such decomposition can be trivially 
increased by one by composing it with an equivalent decomposition 
with a complex conjugate set of coefficients.
We will outline a particular third order scheme, and further 
improved to fourth order, 
which is potentially very useful for practical calculations. We show explicitly that, even though we lose unitarity
of decomposition (in real-time case), the method is asymptotically stable for 
sufficiently small time steps since all the eigenvalues of the decomposition 
remain on the complex unit circle. Even more generally, we show that one gains
an extra order in accuracy and asymptotic stability (independent of the size
of the time step) by renormalizing the state vector after each time step.

We demostrate the accuracy and efficiency of the method by three explicit examples: 
(i) in case of $2\times 2$ matrices the decomposition and its stability can be treated 
analytically, (ii) for exponentials of Gaussian random Hermitean matrices we find that the 
stability threshold (the maximal time-step for which the method is asymptotically stable) 
drops with the inverse power of the dimension of the matrix, and (iii) for a generic 
(non-integrable) interacting spin $1/2$ chain (in one-dimension) we find, surprisingly,
that the stability threshold is independent of the number of spins.

\section{Complex Split-Step Decomposition}
Our main objective is to approximate the exponential operator $U_0=e^{\ii z (A+B)}$, for general bounded
operators $A$ and $B$, and a complex parameter $z$, as a product $U$ of exponential operators
\begin{equation}
\ee^{\ii z(A+B)} = 
\ee^{\ii z p_1 A}
\ee^{\ii z p_2 B}
\ee^{\ii z p_3 A}
\ee^{\ii z p_4 B}
\ee^{\ii z p_5 A} + \mathcal{O}(z^4).
\label{eq:decAB}
\end{equation}
The equations determining the coefficients $\{p_j\}$ that solve the equation above are obtained by expanding
the exponential operators into power series and equating lowest order terms to zero.
It is known that there is no third order ($\mathcal{O}(z^4)$) solution of the five-term ansatz (\ref{eq:decAB}) 
with real coeffients $p_j$. The simplest third order decomposition involves six 
terms \cite{suzukiJPhSocJ61}.
However, allowing the coefficients $p_j$ to be complex, 
there exist {\em two} very simple and symmetric solutions, namely
\footnote{
It was quoted in Ref.\cite{suzukiPLA146} that this solution had already been
proposed by A.D.Bandrauk, however it was claimed in Ref.\cite{bandraukChPhL176} that
the complex coefficient decomposition is unstable and cannot be practically
used for splitting the unitary exponentials, which we show is not precise.}
\begin{equation}
p_1=\overline{p_5}=\frac{1}{4}+\frac{\sqrt{3}}{12}\ii,\quad 
p_2=\overline{p_4}=\frac{1}{2}+\frac{\sqrt{3}}{6} \ii,\quad 
p_3=\frac{1}{2}
\label{eq:coef}
\end{equation}
and the complex conjugate set $\{ \overline{p_j} \}$. 

Let us denote exact exponential as $U_0(z) = \exp(\ii z (A+B))$ and 
third order complex decompositions (C3), given by RHS of (\ref{eq:decAB}) with coefficients (\ref{eq:coef}),
namely $\{p_j\}$, and $\{\overline{p_j}\}$, as $U(z)$ and ${\overline U}(z)$, respectively.
Using some further analysis (which has been performed by means of Mathematica software)
we can show that the next-order-term changes sign when one switches between the two solutions,
namely:
\begin{equation}
U(z) = U_0(z)+K_4 z^4 + \mathcal{O}(z^5) 
\quad \textrm{and} \quad
{\overline U}(z) = U_0(z)-K_4 z^4 + \mathcal{O}(z^5),
\end{equation}
where 
\begin{eqnarray}
K_4 &=& \frac{\ii}{144\sqrt{3}}
((AAAB-BAAA) -3 (AABA-ABAA) - \nonumber \\
&\qquad&\quad\enskip -3 (AABB-BBAA) +6 (ABAB-BABA) + \nonumber \\
&\qquad&\quad\enskip +2 (ABBB-BBBA) +6 (BABB-BBAB) )
\end{eqnarray}
is a Hermitean operator provided that both $A$ and $B$ are Hermitean.

\begin{figure}
\centering
\includegraphics[width=14cm]{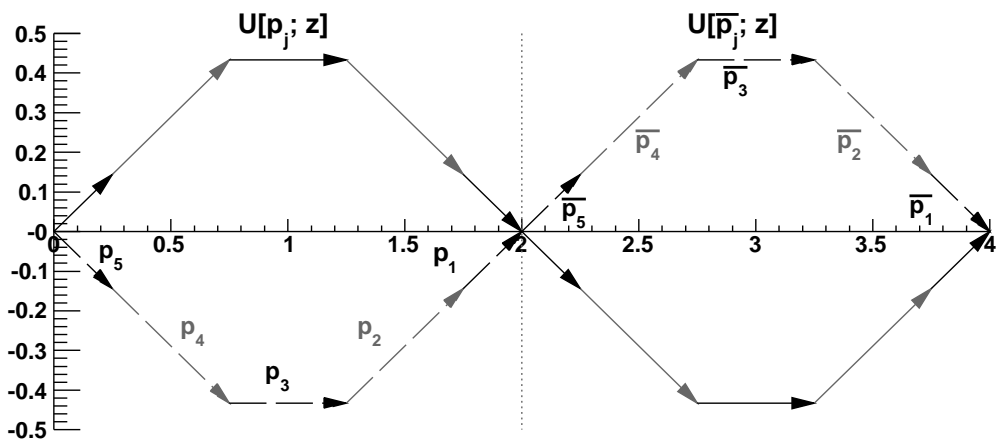}
\caption{Schematic illustration of complex valued split step decomposition. Coefficients $p_j$ can be
considered as shifts in complex time plane, which always move along the real axis.
Two sets of complex coefficients $\{p_i\}$ give a third order decomposition
$\mathcal{O}(z^4)$; their superposition is for an order higher.}
\label{prosenbw}
\end{figure}

Superposition of the two decompositions cancels the $z^4$ term and is therefore
for one order higher, namely of fourth order. 
However, the same, fourth, order can be achieved by alternating both 
decompositions (as illustrated in fig. \ref{prosenbw})
\begin{equation}
{\overline U}(z) U(z) = U_0^2 + (U_0 K_4-K_4 U_0) z^4 + \mathcal{O}(z^5)= U_0^2
+ \mathcal{O}(z^5),
\end{equation}
since $U_0(z)=1+\mathcal{O}(z)$.
Since in usual numerical simulations of exponential operators, for example in quantum time-evolutions, time 
dependent renormalization group methods, or quantum  Monte-Carlo simulations, one needs to make many 
time-steps anyway, the alternation between $U(z)$ and ${\overline U}(z)$ does not represent any practical drawback.

\begin{figure}[!h]
\centering
\includegraphics[width=14cm]{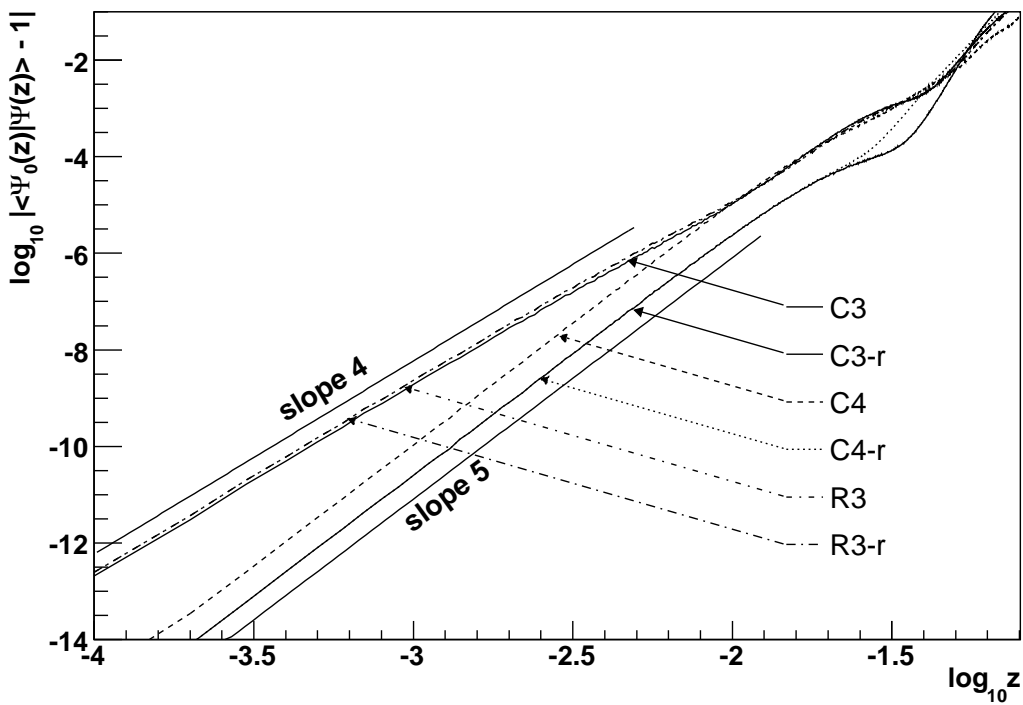}
\caption{An error after two time steps for the third-order real decomposition (R3), 
the third-order complex decomposition (C3), and the fourth-order complex decomposition (C4); the label
'r' denotes renormalization after each time step. As for numerical model we choose $A$ and $B$ to be GUE matrices
of dimension $N=200$ and average the results over 1000 realizations.
}
\label{figfidelity}
\end{figure}

However, we note that with $p_i$ being complex numbers the decomposition $U(z)$ is no longer
{\em strictly unitary} (in the usual case where the operators $A$ and $B$ are Hermitean and the time step $z$ is real) and the time evolved state 
(on which $U$ operates) might explode in norm after a while. 
In order to strictly preserve the norm, the state (vector) may be renormalized at every time step.
One might be afraid that this renormalization would degrade the accuracy of the method. 
However, due to the fact $K_4^\dagger = K_4$ this is not the case, in fact renormalization {\em increases} the accuracy to fourth order
\begin{equation}
\frac{\langle U_0^\dagger(z) U(z)\rangle}{\sqrt{\langle U^\dagger(z) U(z)\rangle}} = 
\frac{1 + \langle K_4\rangle z^4 + \mathcal{O}(z^5)}
         {\sqrt{1+\langle K_4+K_4^\dagger\rangle z^4+\mathcal{O}(z^5)}} = 1 + \mathcal{O}(z^5).
\end{equation}
By $\langle . \rangle := \langle\psi|.|\psi\rangle$ we denote the 
expectation value in some intial state vector $|\psi\rangle$.
In conclusion, the decomposition with one single set of complex coefficients $p_i$ (C3)
is already of the fourth order (${\mathcal O}(z^5)$) if every time step is followed by 
renormalization of the state~(fig. \ref{figfidelity}). As in any application the computational
complexity of performing the sequence of exponential operators on a state vector $U(z)|\psi\rangle$ is
dominating the normalization of the state, this does not represent any drawback of the method.
Still, as we will show later, the method is asymptotically stable, for sufficiently small $z$ even 
without the renormalization.
Figure \ref{figfidelity} shows real numerical errors, in a model in which
$A$ and $B$ are chosen as Gaussian random Hermitean matrices,
after performing two time steps with various decompositions described above 
(using one (C3) or both sets of complex coefficients (C4), and with or without 
renormalization of the state) and compare it with the optimal third order
decomposition with real coefficients (R3).

We can easily generalize our approach to approximate exponentials of three or more noncommuting
bound operators. For example, for three operators, one has nine terms following a sequence
$ABCBABCBA$ which is obtained from $ABABA$ (\ref{eq:decAB}) by replacing each inner operator $B$ by $BCB$ 
(and dividing the coefficient in front of $B$ by two)
\begin{equation}
\ee^{\ii z(A+B+C)} = 
\ee^{\ii z p_1 A}
\ee^{\ii z p_1 B}
\ee^{\ii z p_2 C}
\ee^{\ii z p_1 B}
\ee^{\ii z p_3 A} 
\ee^{\ii z p_4 B} 
\ee^{\ii z p_5 C} 
\ee^{\ii z p_4 B} 
\ee^{\ii z p_5 A} 
+ \mathcal{O}(z^4)
\end{equation}
and using the same set of coefficients (\ref{eq:coef}), 
or its complex conjugate.
Generally, a formula for a sum of $n$ operators involves $4n-3$ terms
\begin{eqnarray}
&& \exp\left(\ii z(A_1+\ldots A_n)\right) = \nonumber \\
&&
\ee^{\ii z p_1 A_1}
\ee^{\ii z p_1 A_2}
\cdots
\ee^{\ii z p_1 A_{n-1}}
\ee^{\ii z p_2 A_{n}}
\ee^{\ii z p_1 A_{n-1}}
\cdots
\ee^{\ii z p_1 A_2}
\times \nonumber \\
&&
\ee^{\ii z p_3 A_1} 
\ee^{\ii z p_5 A_2}
\cdots
\ee^{\ii z p_5 A_{n-1}}
\ee^{\ii z p_4 A_{n}}
\ee^{\ii z p_5 A_{n-1}}
\cdots
\ee^{\ii z p_5 A_2}
\ee^{\ii z p_5 A_1}.
\label{eq:general}
\end{eqnarray}
It is interesting to note that the general optimal third order solution with real coefficients (R3)
uses just one term more for the case $n=2$, namely six, whereas for general $n$ case it needs 
$5 n - 4$ terms, which is $n-1$ terms more than the complex solution above (\ref{eq:general}).

As we have mentioned before, without the renormalization complexness of the coefficients may cause the 
exponential instability of the method. However, it turns out that the decomposition is absolutely 
stable for small enough steps $z$. The reason for such an interesting behaviour is that the
eigenvalues of the operator $U(z)$ lie all on complex unit circle for sufficiently small $z$,
and this property grants the asymptotic stability even if $U(z)$ is not exactly unitary.
There is typically a threshold, i.e. a critical value of $z_{\rm max}$ such that at $z=z_{\rm max}$ 
two eigenvalues of $U(z)$ collide and leave the unit circle and then the method ceases to be 
asymptotically stable. Such a behaviour can be explicitly proven for operators chosen from the 
space of $2\times 2$ matrices (see the following section) and is conjectured in general.

\section{Examples}

\begin{figure}
\centering
\includegraphics[width=14cm]{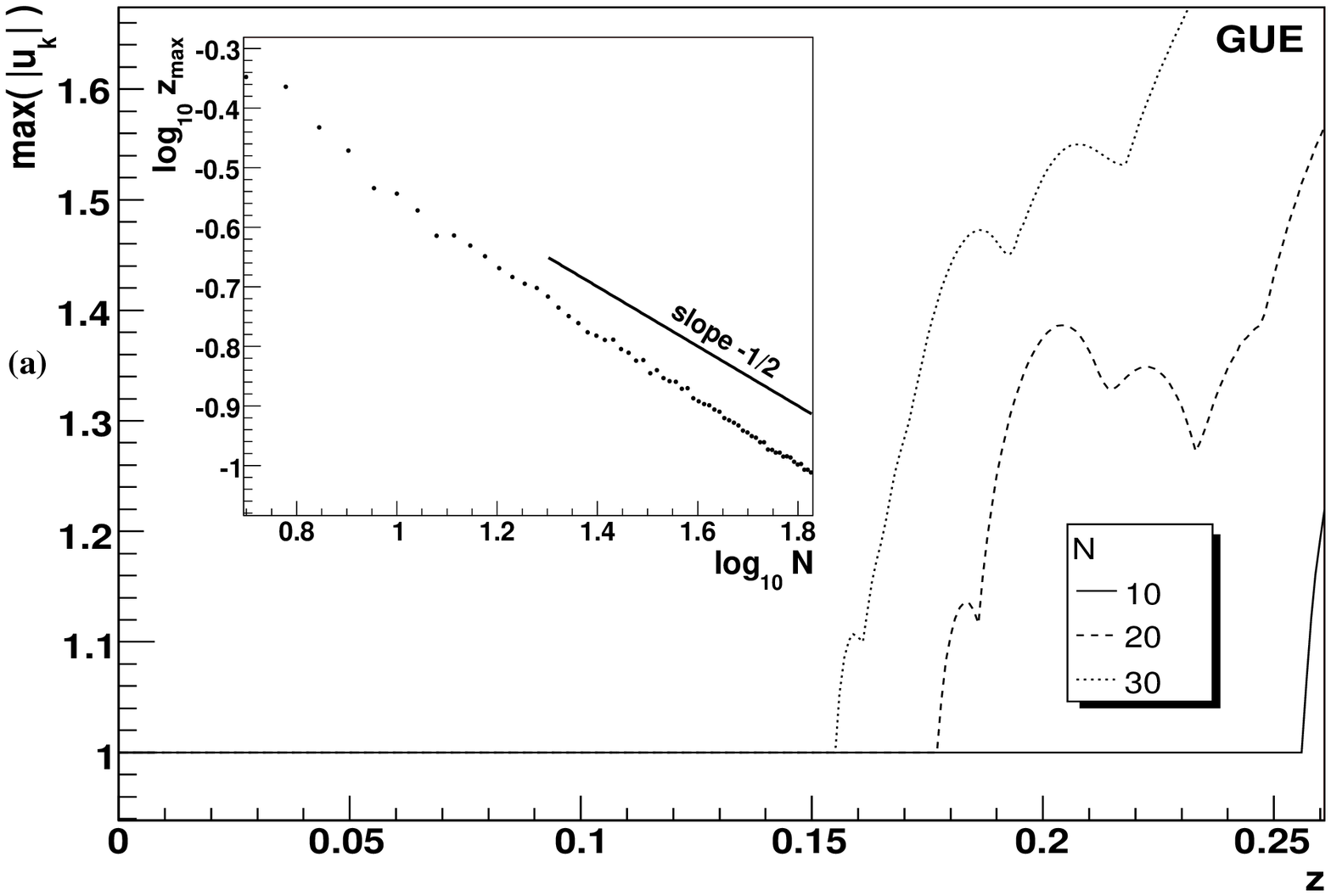} \\
\includegraphics[width=14cm]{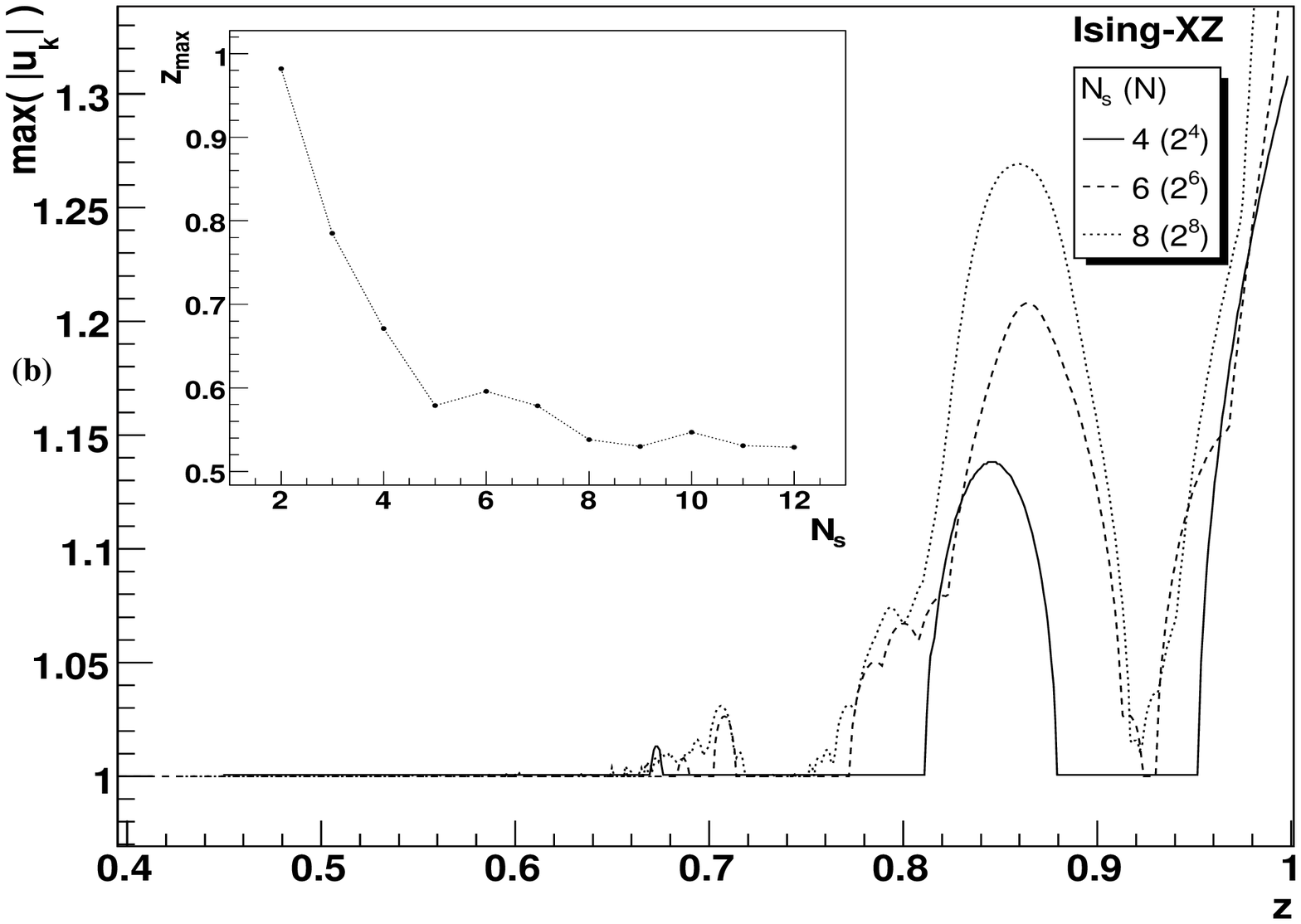}
\caption{Maximal size of the eigenvalue of the approximate evolution operator $U(z)$.
The upper plot (a) shows the case of GUE matrices while the lower plot (b) shows the
case of Ising spin chain in tilted magnetic field (see text). Different curves refer to
systems of different sizes (b), or different matrix dimensions (a). The insets show critical 
threshold $z_{\rm max}$ as a function of the system size/matrix dimension.}
\label{figmaxz}
\end{figure}

First, let us consider a numerical example of calculating the exponential of 
$H=A+B$ where $A$ and $B$ are Gaussian random Hermitean matrices chosen at 
random from the Gaussian Unitary Ensemble \cite{mehtabook}. 
Figure~\ref{figmaxz}a shows that the maximal size of eigenvalue of $U(z)$ 
is exactly equal to one until some point described by the threshold step size 
$z_{\rm max}$. Numerical results suggest the following dependence of the threshold on the Hilbert
space dimension $N$, $z_{\rm max} \propto 1/N^\alpha$, with $\alpha \approx 0.5$,
which we believe is the worst case scenario for generic systems.
\\\\
As a second example, we consider a non-trivial physical model where the 
matrices of operators $A$ and $B$ are very 
sparse and thus far from the full random matrix model, namely we 
consider time evolution in the quantum Ising spin 1/2 chain in a 
tilted homogeneous magnetic field (e.g. recently considered in the context of
heat transport \cite{mejia}) described by the hamiltonian 
$H=\sum_{n=1}^N\left\{-J \sigma_{n}^z\sigma_{n+1}^z + g_x \sigma_n^x + g_z \sigma_n^z\right\}$.
Here, $\sigma_n^{x,y,z}$, $n=1\ldots N_s$, represent a set of independent Pauli matrices.
In figure~\ref{figmaxz}b we show a very interesting result for this model 
(in particular, for the parameter values 
$J=1$, $g_x=0.4$, $g_z=0.8$
which lie in  the so-called ``quantum chaotic'' regime \cite{mejia}), 
namely that the threshold step size $z_{\rm max}$ is asymptotically independent of 
the size $N=2^{N_s}$ of the system. We conjecture that this is in general true for numerical simulations 
of finite (spin) quantum systems with local interaction, and for such our method of simulation 
of time-evolution should be very roboust.
\\\\
As for the last example, we make analytical consideration of the simplest case where our operators can be 
represented by $2\times 2$ matrices. In order to understand the transition in the stability 
(collision of eigenvalues of $U(z)$ on the unit circle) one can generally parametrize the operators 
$A$ and $B$ by Pauli operators $\sigma^j$, $j=1,2,3$,
\begin{equation}
A = a_0 1 + \sum_{j=1}^3 a_j \sigma^j
\quad\textrm{and}\quad
B = b_0 1 + \sum_{j=1}^3 b_j \sigma^j.
\end{equation}
The coefficients $\{a_j\}$, and $\{b_j\}$ 
are all real since matrices $A$ and $B$ are Hermitean, and furthermore
matrices $A$ and $B$ can always be chosen traceless by setting $a_0=b_0=0$ 
without losing generality.
It is obvious that, since $\det U = e^{\ii z \textrm{Tr} H}$, where $H=A+B$, that
decomposition (\ref{eq:decAB}) for two $2\times 2$ matrices can also be expressed
in terms of Pauli matrices and some coefficients $\{g_j\}$.
Using the ansatz (\ref{eq:decAB}) we write
\begin{equation}
\ee^{\ii z p_1 \sum_j a_j \sigma^i}
\ee^{\ii z p_2 \sum_j b_j \sigma^i}
\ee^{\ii z p_3 \sum_j a_j \sigma^i}
\ee^{\ii z p_4 \sum_j b_j \sigma^i}
\ee^{\ii z p_5 \sum_j a_j \sigma^i}
= 
\ee^{\ii z \sum_j g_j \sigma^j}.
\end{equation}
Of course, $g_j$ are no longer real in general.
Eigenvalues of the operator 
$U(z)=e^{\ii z \sum_j g_j \sigma^j}$
are $e^{\pm \ii z \sqrt{\sum_j g_j^2}}$
which gives the condition for the asymptotic stability:
namely the number $\gamma^2=\sum_j g_j^2$ should be {\em real} and {\em positive}, 
$\gamma^2 \in \mathbf{R}^+$. In order to simplify the notation, let us take 
$\gamma = +\sqrt{\sum_j g_j^2}$, and similarly write
$\alpha=\sqrt{\sum_j a_j^2}, \beta=\sqrt{\sum_j b_j^2}$, 
and introduce normalized coefficients $\gamma_i = g_i/\gamma, 
\alpha_j = a_j/\alpha,
\beta_j = b_j/\beta$.
The condition for asymptotic stability now simply reads $\gamma \in \mathbf{R}$.
Using straightforward calculation $\gamma$ can be expressed as 
$\gamma = \frac{1}{z}\arccos(\frac{1}{2}\textrm{Tr}\,e^{\ii z \sum_i g_i \sigma_i})$
and is, interestingly, 
only a function of the magnitudes $\alpha$, $\beta$ and z-projections $\alpha_3$ and $\beta_3$:
\begin{eqnarray}
\gamma(z) &=& 
\frac{1}{z}\arccos Q(z),
\quad \textrm{where}\nonumber \\
Q(z) &=& 
\frac{1}{8} 
\Bigg(
        \Big(1-\alpha_3^2+(1+3 \alpha_3^2)\cos(\alpha z)\Big)
			\Big((1+\beta_3^2)\cos(\beta z) + \nonumber \\
		&+& (1-\beta_3^2)\cosh(\frac{\beta z}{\sqrt{3}})\Big)
			-2 \alpha_3 (3+\alpha_3^2)\beta_3\sin(\alpha z)\sin(\beta z) +\nonumber\\
        &+& 2(1-\alpha_3^2)\cosh(\frac{\alpha z}{2\sqrt{3}}) 
			\Big((1+\beta_3^2)\cos(\frac{\alpha z}{2})\cos(\beta z) + \nonumber\\
        &+&(1-\beta_3^2)\cos(\frac{\alpha z}{2})\cosh(\frac{\beta z}{\sqrt{3}})
			- 2\alpha_3\beta_3\sin(\frac{\alpha z}{2})\sin(\beta z)
        \Big)
\Bigg)
\label{eq:gamma}
\end{eqnarray}
Now the stability condition reduces to $\left|Q(z)\right|\leq 1$.
For small steps $z$ the expression $Q(z)$ in (\ref{eq:gamma}) 
can be written as a power series in $z$
\begin{equation}
Q(z) =
1 - \frac{1}{6}(\alpha^2 + \beta^2 + 2 a_3^2 + 2 b_3^2 + 6 a_3 b_3) z^2 + \mathcal{O}(z^4).
\end{equation}
It can easily be proven diagonalizing the quadratic form that 
the $z^2$ term is always nonpositive, hence the decomposition scheme indeed is always 
(for any $a_j,b_j$)
stable, for small steps $z$.

\begin{figure}
\centering
\includegraphics[width=9cm]{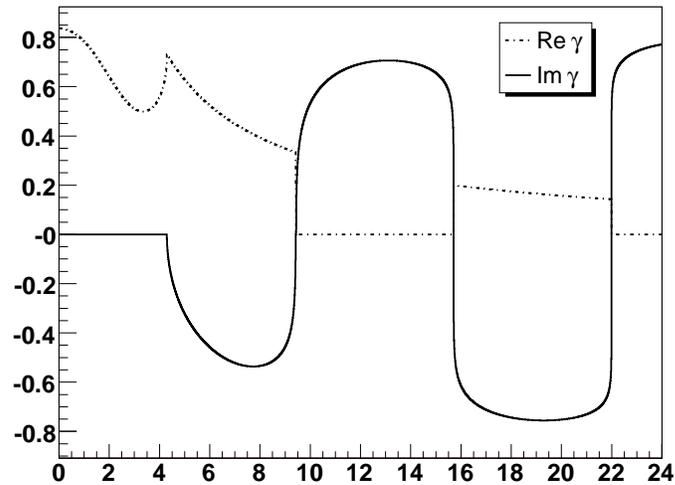}
\caption{Illustration of the stability threshold for $2\times 2$ case. Since matrices are traceless, collision of 
eigenvalues of $U(z)$
takes place on the real axis. In the figure we plot ${\rm Re} \gamma$ (dashed) and ${\rm Im} \gamma$
(full), as a function of $z$ for the case $\alpha=\beta=1$ and $\alpha_3=\beta_3=0.1$. }
\label{figpauli}
\end{figure}

Figure~\ref{figpauli} illustrates how eigenvalues for small 
steps $z$ always lie on the unit circle in
the complex plane. When the step $z$ is being increased, the
eigenvalues are travelling along the unit circle, one in clockwise
and the other in the counter-clockwise direction. 
At some point, namely at $z=z_{\rm max}$, a collision occurs and a pair of eigenvalues bounce off the unit circle - then
$\gamma$ becomes complex. However, because of the
restriction $\left|\det U\right|=1$ their product remains on the unit circle.
Our $2\times 2$ matrices $A$ and $B$ are assumed to be traceless therefore collisions always 
occur on the real axis and eigenvalues are both real during the bounce.

\section{Conclusion}
We have proposed a simplex explicit complex-coefficient split-step 
decomposition of an operator exponential, based on 
Suzuki's scheme, for a sum of arbitrary number of
operators.
As compared to an optimal scheme with real coefficients our 
scheme requires less terms for the same order, furthermore we can gain an extra order
at no additional expense. Despite having complex coefficients the decomposition is
always stable for sufficiently small step size, and can be stablilized by additional
renormalization of the state vector.

We suggest that our method may be used in conjunction with other methods for efficient time
evolution of complex quantum systems (one application has already been done in Ref.\cite{mejia}),
or interacting many body quantum systems, like for example with time-dependent DMRG methods \cite{white,vidal} 
where efficient and accurate estimation of operator exponentials for short time steps 
is one of the cruicial black-box operations.

\section*{Acknowledgements}

We acknowledge support by Slovenian Research Agency, in particular from the grant J1-7347 and the programme P1-0044.

\end{document}